\documentclass[twocolumn]{aastex62}
\usepackage{amsmath}
\usepackage{lineno}

\newcommand{\feii}{\mbox{Fe{\small II}}}
\newcommand{\mgii}{\mbox{Mg{\small II}}}

\graphicspath{{./}{figures/}}

\received{...}
\revised{...}
\accepted{...}

\submitjournal{ApJ}

\shorttitle{Disk Composition and the Evolution of Embedded Stars}
\shortauthors{Dittmann, Jermyn, \& Cantiello}

\begin{document}
\title{The Influence of Disk Composition on the Evolution of Stars in the Disks of Active Galactic Nuclei}

\correspondingauthor{Alexander J. Dittmann}
\email{dittmann@astro.umd.edu}

\author[0000-0001-6157-6722]{Alexander J.~Dittmann}
\affil{Department of Astronomy and Joint Space-Science Institute, University of Maryland, College Park, MD 20742-2421, USA}
\affil{Theoretical Division, Los Alamos National Laboratory, Los Alamos, NM 87545, USA}

\author[0000-0001-5048-9973]{Adam S. Jermyn}
\affil{Center for Computational Astrophysics, Flatiron Institute, 162 5th Avenue, New York, NY 10010, USA}

\author[0000-0002-8171-8596]{Matteo Cantiello}
\affil{Center for Computational Astrophysics, Flatiron Institute, 162 5th Avenue, New York, NY 10010, USA}
\affil{Department of Astrophysical Sciences, Princeton University, Princeton, NJ 08544, USA}

\begin{abstract}
Disks of gas accreting onto supermassive black holes, powering active galactic nuclei (AGN), can capture stars from nuclear star clusters or form stars in situ via gravitational instability. The density and thermal conditions of these disks can result in rapid accretion onto embedded stars, dramatically altering their evolution in comparison to stars in the interstellar medium. Theoretical models predict that, when subjected to sufficiently rapid accretion, fresh gas replenishes hydrogen in the cores of these stars as quickly as it is burned into helium, reaching a quasi-steady state. Such massive, long-lived (``immortal'') stars may be capable of dramatically enriching AGN disks with helium, and would increase the helium abundance in AGN broad-line regions relative to that in the corresponding narrow-line regions and hosts. We investigate how the helium abundance of AGN disks alters the evolution of stars embedded therein. We find, in agreement with analytical arguments, that stars at a given mass are more luminous at higher helium mass fractions, and so undergo more radiation-driven mass-loss. We further find that embedded stars tend to be less massive in disks with higher helium mass fractions, and that immortal stars are less common in such disks. Thus, disk composition can alter the rates of electromagnetic and gravitational wave transients as well as further chemical enrichment by embedded stars.
\end{abstract}


\section{Introduction} \label{sec:intro}
Active galactic nuclei (AGN) are powered by the accretion of massive gas disks onto supermassive black holes \citep{1969Natur.223..690L,2008ARA&A..46..475H}. The outer regions of these disks can become gravitationally unstable, leading to star formation \citep[e.g.][]{1980SvAL....6..357K,2003MNRAS.339..937G,2020MNRAS.493.3732D,2022arXiv220510382D}. 
Furthermore, stars can be captured by the disk from nuclear star clusters, which are ubiquitous at least in quiescent galaxies \citep[e.g.][]{2020A&ARv..28....4N}, through a variety of mechanisms.
Gas torques may capture initially misaligned stars into the disk \citep[e.g.][]{1993ApJ...409..592A,1995MNRAS.275..628R}, further align stellar orbits the midplane of the disk \citep[e.g.][]{2004ApJ...602..388T}, and circularize initially eccentric stellar orbits \citep[e.g.][]{1991MNRAS.250..505S,2020ApJ...889...94M}.

Regardless of the mechanisms by which stars become embedded in AGN disks, accretion from the disk can profoundly alter their evolution \citep{2021ApJ...910...94C,2021ApJ...916...48D,2021ApJ...914..105J}, leading to the formation of very massive stars \citep[e.g.][]{2021ApJ...910...94C,2021ApJ...911L..14W}, gamma-ray bursts \citep[e.g.][]{2021ApJ...906L...7P,2021ApJ...914..105J,2021ApJ...911L..19Z,2021MNRAS.508.1842D}, and gravitational wave event progenitors \citep[e.g.][]{2017MNRAS.464..946S,2022ApJ...929..133J,2022arXiv220510382D}. Evolutionary models which embed stars in a medium of constant density and temperature predict that stars may reach a quasi-steady state, undergoing no chemical evolution \citep{2021ApJ...910...94C,2021ApJ...916...48D}. Stars reach such a state (``immortal'') when the accretion of fresh gas can supply hydrogen to stellar cores faster than it is spent powering the star, leading to a balance between accretion and radiation-driven mass loss. 

Immortal stars may, in principle, be able to survive as long as the AGN disk persists. Observations indicate that individual AGN accretion episodes may be as short as $\sim10^5$ years \citep{2015MNRAS.451.2517S,10.1093/mnrasl/slv098}, although galactic nuclei are expected to be active for total durations on the order of $\sim10^8$ years \citep[e.g.][]{2004cbhg.symp..169M}. Depending on the accretion rate through a given disk, a population of immortal stars may be able to convert almost all of the hydrogen accreting through the disk into helium \citep{2022ApJ...929..133J}. However, previous calculation of stellar evolution in AGN disks assumed a fixed disk helium mass fraction of $Y\equiv m_{\rm He}/m_{\rm gas}=0.28$, which limits their applicability if AGN disks are indeed enriched in helium.

A few studies have used spectroscopy to constrain the relative abundance of helium to hydrogen $(y\equiv n_{\rm He}/n_{\rm H})$ in active galaxies. For example, \citet{1971ApJ...163..235B} estimated $y$ in the range of $0.003-0.2$ in a sample of 14 quasars, and \citet{1975ApJ...201...26B} estimated $y$ in the range of $\sim 0.1$ to $\sim 0.3$ in a sample of 14 quasars at redshifts $z\approx 0.2$. Naively estimating $Y\approx4y/(1+4y)$, these measurements suggest values of $Y$ ranging from $\sim0.012$ to $\sim0.54$. However, these measurements may underestimate the helium abundance in these sources for reasons such as the steepness of the ionizing spectrum \citep{1971ApJ...167L..27W} or variability of the ionizing flux over time \citep{1973ApJ...181..627J}. Furthermore, modeling sufficiently reliable to determine the helium abundance from helium-to-hydrogen line ratios has remained elusive at the densities of quasar broad line regions due to factors such as the importance of three-body recombination \citep[e.g][]{1990agn..conf...57N}. 

Recently \citet{2022MNRAS.tmp.1674D} measured $Y$ in a sample of 65 Seyfert 2 narrow-line regions at redshifts $z\lesssim0.2$, finding values of $Y$ ranging from $\sim0.2$ to $\sim0.46$. In contrast, the sample of 85 star forming regions studied in \citet{2022MNRAS.tmp.1674D} using the same methodology range in $Y$ from $\sim0.18$ to $\sim0.3$, suggesting significant helium enrichment in active galaxies. 

Stars may also enrich AGN disks with metals, at least when those stars are able to reach later stages of nuclear burning, especially if those stars end their lives as supernovae \citep[e.g.][]{1993ApJ...409..592A,2021ApJ...916...48D}. Spectroscopic measurements suggest that quasar broad line regions have metallicities at least a few times the solar value with negligible evolution over cosmic time \citep[e.g.][]{2009A&A...494L..25J,2018MNRAS.480..345X,2020ApJ...898..105O,2022ApJ...925..121W,2022MNRAS.513.1801L,2022arXiv220802387G}. Additionally, studies have found that AGN broad-line region metallicity correlates with supermassive black hole (SMBH) mass, suggesting that chemical enrichment is tied to SMBH growth \citep[e.g.][]{2009A&A...503..721M,2018MNRAS.480..345X,2022ApJ...925..121W}. More specifically, the redshift-independence of the broad-line region \feii/\mgii \,flux ratio \citep[e.g.][]{2003ApJ...596..817D,2017ApJ...849...91M,2020ApJ...898..105O,2021ApJ...923..262Y}, for which overproduction of iron and large [Fe/Mg] abundance ratios are the likely origins \citep{2017ApJ...834..203S,2020ApJ...904..162S}, can be explained by supernovae produced by stars in AGN disks that were made more massive through accretion \citep[e.g.][]{1993ApJ...409..592A,2022MNRAS.512.2573T}. Analyses of the iron K$\alpha$ X-ray fluorescent line have also suggested that the iron abundance in the inner regions of AGN disks may be significantly supersolar \citep[e.g.][]{1995Natur.375..659T,1997ApJ...488L..91N}, although these inferences are subject to large systematic uncertainties \citep{2018ASPC..515..282G,2018ApJ...855....3T}. Understanding which disk conditions, including sound speed ($c_s$), density ($\rho$), and helium mass fraction, lead to stars ending their lives via supernova, is crucial to understanding the capacity of stars to enrich AGN disks with metals. 

Here we present a survey of stellar evolution in AGN disks focusing on the role of the disk helium abundance. In Section \ref{sec:analytic} we review a number of analytic arguments suggesting that as the disk $Y$ increases, stars should be more luminous at a given mass, leading to higher densities being required for immortal stars to form, and stars generally being less massive at constant densities and sound speeds. In Section \ref{sec:numerical} we briefly review the numerical methods which we use to model stellar evolution in AGN disks, and we present the results of our simulations in Section \ref{sec:results}. We discuss our results in the context of various observations in Section \ref{sec:discussion} and conclude in Section \ref{sec:conclusion}.

\section{Analytic Considerations}\label{sec:analytic}
First and foremost, the AGN disk modifies the evolution of embedded stars by providing an ample supply of dense gas. If the relative velocity between a given star and the ambient medium is low, the accretion rate onto the star can be approximated by the Bondi rate \citep{1952MNRAS.112..195B}
\begin{equation}\label{eq:bondi}
\dot{M}_B=\eta \pi R_B^2\rho c_s,
\end{equation}
where $\rho$ and $c_s$ are respectively the density and sound speed of the ambient medium, $\eta\lesssim1$ is an efficiency factor, and $R_B$ is the Bondi radius given by
\begin{equation}
R_B=\frac{2GM_*}{c_s^2},
\end{equation}
where $G$ is the gravitational constant and $M_*$ is the mass of the accreting star.

The accretion rate onto AGN stars can be substantially reduced under a range of conditions, such as the presence of a large relative velocity ($\Delta v \gtrsim c_s$) between the star and ambient medium \citep[][e.g., for stars on retrograde orbits]{1939PCPS...35..592H}. The accretion rate may also be reduced when the Bondi radius would exceed $H$, the pressure scaleheight of the AGN disk, or the Hill radius $R_H=r_*(GM_*/M_\bullet)^{1/3}$ \citep[][where $r_*$ is the distance between the star and SMBH, and $M_\bullet$ is the mass of the SMBH]{hill1878researches}, the distance from the star where gas becomes bound to it rather than the SMBH \citep[see, e.g.][]{2007ApJ...660..791D,2020MNRAS.498.2054R}.
Herein we use just the Bondi accretion rate above to focus on the effects of disk composition, but note that \citet{2021ApJ...916...48D} previously studied the influence on the evolution of AGN stars of various physical effects which reduce the accretion rate compared to the Bondi rate.

As stars accrue mass from the disk, their luminosities ($L_*$) increase rapidly, $L_*\propto M_*^3$ \citep[e.g.][for gas-pressure dominated stars]{1992isa3.book.....B}. Eventually the momentum carried by this radiation can compete with the gravity of the star in governing the motion of gas, the two accelerations balancing one another when $L_*\kappa/4\pi r^2c=GM_*/r^2$, where $c$ is the speed of light, $\kappa$ is the opacity of the gas, and $r$ is the distance between the star and gas parcel in question. The competition of these accelerations defines the Eddington luminosity
\begin{equation}
L_{\rm Edd}=4\pi G M_*c/\kappa.
\end{equation}
As $L_*$ approaches $L_{\rm Edd}$, the accretion rate onto the star should decrease, potentially to zero. However, deviations from spherical symmetry, such as the greater luminosity of rotating stars near their poles than their equators \citep[e.g][]{1924MNRAS..84..665V,1967ZA.....65...89L}, or from density asymmetries in the ambient medium and the instabilities which ensue upon their irradiation \citep[e.g][]{2013MNRAS.434.2329K,2014ApJ...796..107D}, may prevent accretion from being entirely halted. Accordingly, we employ the phenomenological prescription\footnote{See \citet{2021ApJ...916...48D} for an exploration of alternative models of accretion rate reduction due to radiative feedback.} \citep{2021ApJ...910...94C}
\begin{equation}
\dot{M}_{\rm acc}=\dot{M}_B\left(1-\tanh{\left[\frac{L_*}{L_{\rm Edd}}\right]}\right).
\end{equation}

Following \citet{2021ApJ...910...94C}, we approximate that as $L_*$ approaches $L_{\rm Edd},$ near- or super-Eddington continuum-driven winds dominate stellar mass loss. Along the lines of previous work \citep[e.g.][]{1986ApJ...302..519P,2011ApJS..192....3P,2020ApJ...904L..13R}, we assume an outflow at the escape velocity from the surface $v_{\rm esc}=(2GM_*/R_*)^{1/2}$ and an associated phenomenological mass loss rate\footnote{See \citet{2021ApJ...914..105J} for an exploration of enhanced mass loss driven by stellar rotation.}
\begin{equation}
\dot{M}_{\rm loss}=-\frac{L_*}{v_{\rm esc}^2}\left[1 + \tanh\left(\frac{L_*-L_{\rm Edd}}{0.1L_{\rm Edd}}\right)\right].
\end{equation}

As embedded stars accrete from the AGN disk, their composition becomes strongly influenced by that of the surrounding gas. Ignoring metallicity, the mean molecular weight ($\mu$) of a fully ionized gas is given by $\mu=(2-5Y/4)^{-1},$ such that a higher helium mass fraction in the disk, and hence in the star, leads to a higher stellar mean molecular weight. At a fixed stellar mass, $L_\star$  increases with mean molecular weight, e.g. $L_*\propto \mu^4 M^3$ for a simplified stellar model with uniform molecular weight and constant opacity \citep[][chapter 20]{2013sse..book.....K}. Thus, AGN stars in higher-$Y$ disks will tend to be more luminous at a given mass, potentially driving stronger outflows and halting accretion at lower masses.

However, changes in the composition of AGN stars and the disk can change gas opacities and thus the Eddington luminosity, as impingement of radiation upon more opaque gas can more easily overcome the pull of gravity. Considering just electron scattering opacity, $\kappa\approx0.2(2-Y)\,\rm{cm^2g^{-1}}$. Thus, increases in $Y$ can decrease gas opacity enough to reduce $L_{\rm Edd}$ by up to a factor of 2 compared to the value corresponding to pure hydrogen gas. In more helium-rich AGN disks, then, it may be more difficult for stars to drive outflows and stave off accretion. However, the electron scattering opacity may also severely underestimate the actual opacity due to features such as the iron opacity peak, or the potentially stronger opacity peaks due to helium recombination  \citep[e.g.][]{2015ApJ...813...74J,2016ApJ...827...10J,2018Natur.561..498J,2019ApJ...883..106C,2022arXiv220600011J}, so we present results using both opacity formulas and caution that there is still uncertainty in the opacities relevant to mass loss and the accretion stream, even in this already simplified gray opacity treatment.

\subsection{Timescales}
A first-order picture of the evolution of AGN stars can be gleaned from the balance between the main sequence timescale of such a star in the absence of accretion and the typical accretion timescale onto that star due to Bondi accretion.
The latter can be characterized by the initial mass-doubling timescale of the star,\footnote{The mass-doubling timescale, due to its dependence on mass for Bondi accretion, is also half of the time required for runaway accretion in the absence of feedback.} which for accretion at the Bondi rate is given by
\begin{equation}
\tau_2=\frac{c_s^3}{8\pi\eta G^2\rho M_0},
\end{equation}
where $M_0$ is the initial mass of the star, or in terms of a characteristic density ($\rho_0\equiv10^{-18}\,\rm{g\,cm^{-3}}$) and sound speed $(c_{s,0}\equiv10^6\,\rm{cm\,s^{-1}})$ appropriate to the outer gravitationally unstable (or marginally stable) regions of AGN disks \citep[e.g.][]{2003MNRAS.341..501S,2021ApJ...910...94C}, 
\begin{equation}
\tau_2\approx2.2\times10^8\,\rm{yr} \left(\frac{\rho}{\rho_0}\right)^{-1}\left(\frac{c_s}{c_{s,0}}\right)^3\left(\frac{M_0}{M_\odot}\right)^{-1}.
\end{equation}

\begin{figure}
\includegraphics[width=\linewidth]{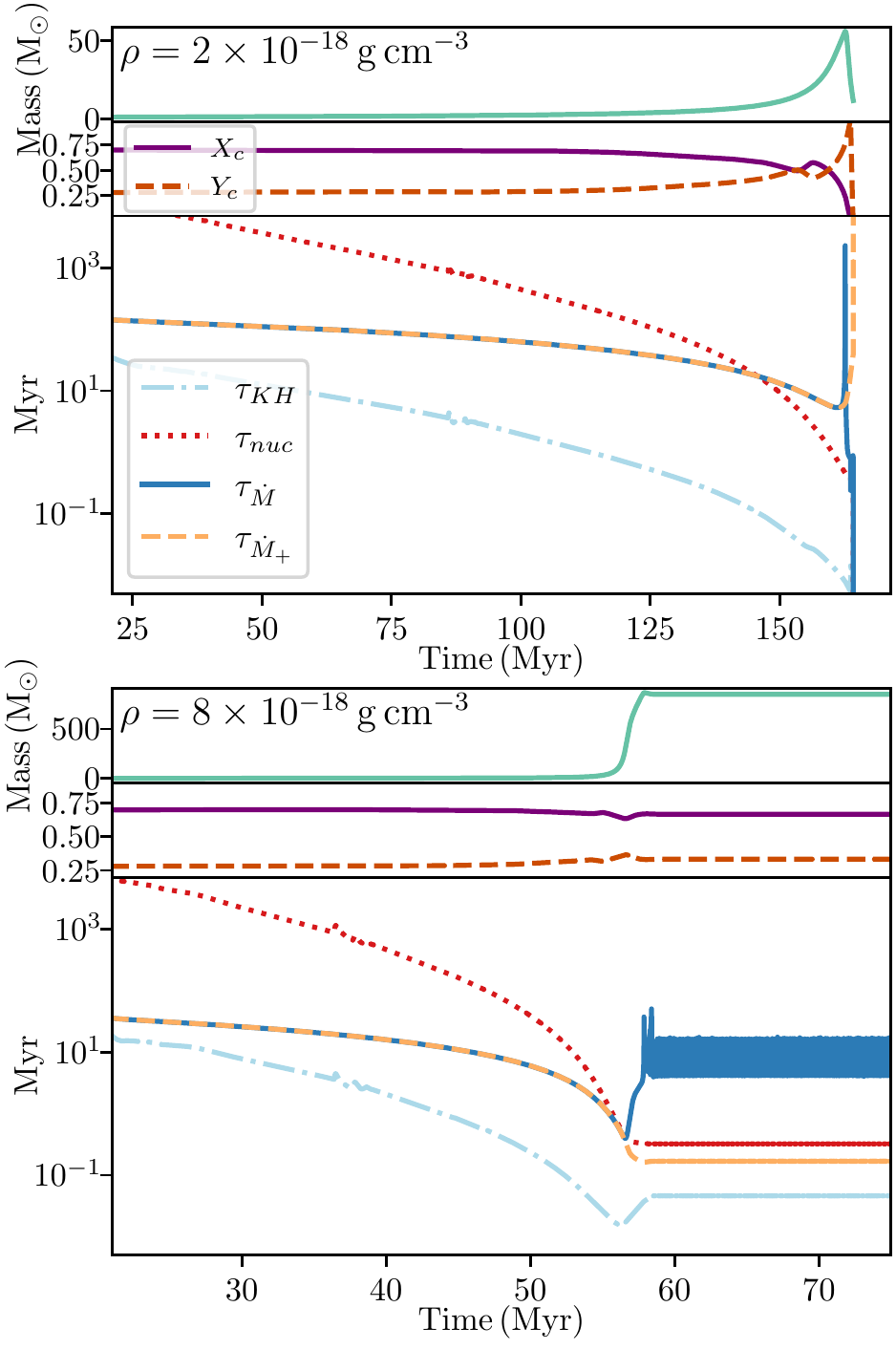}
\caption{The evolution of fiducial intermediate (top panel trio, using a density of $\rho=2\times10^{-18}\,\rm{g\,cm^{-3}}$) and immortal (bottom panel trio, using a density of $\rho=8\times10^{-18}\,\rm{g\,cm^{-3}}$) stars. In each trio of panels, the upper plot shows the stellar mass over time, the middle plot shows the core hydrogen and helium mass fractions over time, and the lower plot shows the evolution of various timescales over time. All simulations used an ambient sound speed of $c_s=10^6\,\rm{cm\,s^{-1}}$. In both cases, the balance between the nuclear and accretion timescales determines the fate of the star, which hinges on whether or not the former becomes shorter than the latter.}
\label{fig:timescales}
\end{figure}

If $\tau_2$ is longer than the time spent by a given star on the main sequence, then the evolution of that star is minimally altered.\footnote{However, see \citet{1989MNRAS.238..427T} for an investigation of the effects of irradiation by AGNs on the evolution of non-embedded stars.} However, if the accretion timescale is shorter than the main sequence lifetime of a star, its structure can be altered considerably. Ultimate stellar fates depend on whether mass loss is able to overcome or balance accretion. \citet{2021ApJ...910...94C} argues that when the nuclear burning timescale ($\tau_{\rm nuc}$) is shorter than, but comparable to, the accretion timescale, stars may accrete substantial mass before approaching the Eddington luminosity, driving extreme mass loss, and returning to moderate masses of $\sim10\,M_\odot$. We refer to such stellar models as `intermediate,' and the evolution of such a model is displayed in the upper panels of Figure \ref{fig:timescales}. 

The upper panels of Figure \ref{fig:timescales} illustrate the balance of various timescales more quantitatively over the course of the evolution of an intermediate star. Here, we plot a rough estimate of the nuclear burning timescale $\tau_{\rm nuc}\sim10^{10}(M/M_\odot)(L/L_\odot)^{-1}\,\rm{yr}$, the Kelvin-Helmholtz timescale $\tau_{\rm KH}\sim0.75GM_*^2R_*^{-1}L_*^{-1},$ the overall timescale for mass changes $\tau_{\dot{M}}\equiv M_*/|\dot{M}_{\rm acc}-\dot{M}_{\rm loss}|$, and the timescale for mass change due to accretion $\tau_{\dot{M}_+}\equiv M_*/\dot{M}_{\rm acc}$. For the intermediate stellar model, the accretion timescale is initially shorter than the nuclear timescale. However, as the star grows more massive and luminous, the nuclear timescale shrinks more quickly than the accretion timescale, eventually becoming shorter. The star then becomes sufficiently luminous to stave off further accretion almost entirely, in this case ejecting roughly forty solar masses of material.  

On the other hand, if the accretion timescale is sufficiently short, accretion and mass loss can balance one another, resulting in almost fully-convective stellar models with masses which are (on average) constant in time. Such stars are able to efficiently mix accreted material from the surface into their cores, and thus enrich their outer layers and winds with fusion byproducts, potentially enriching the disk with helium \citep{2021ApJ...916...48D,2022ApJ...929..133J}. We refer to such stellar models as ``immortal,'' and the evolution of such a model is shown in the bottom panels of Figure \ref{fig:timescales}. In this case, although the nuclear timescale becomes much shorter as the star accretes, it is unable to outpace the shrinking of the accretion timescale.

\begin{figure*}
\includegraphics[width=\linewidth]{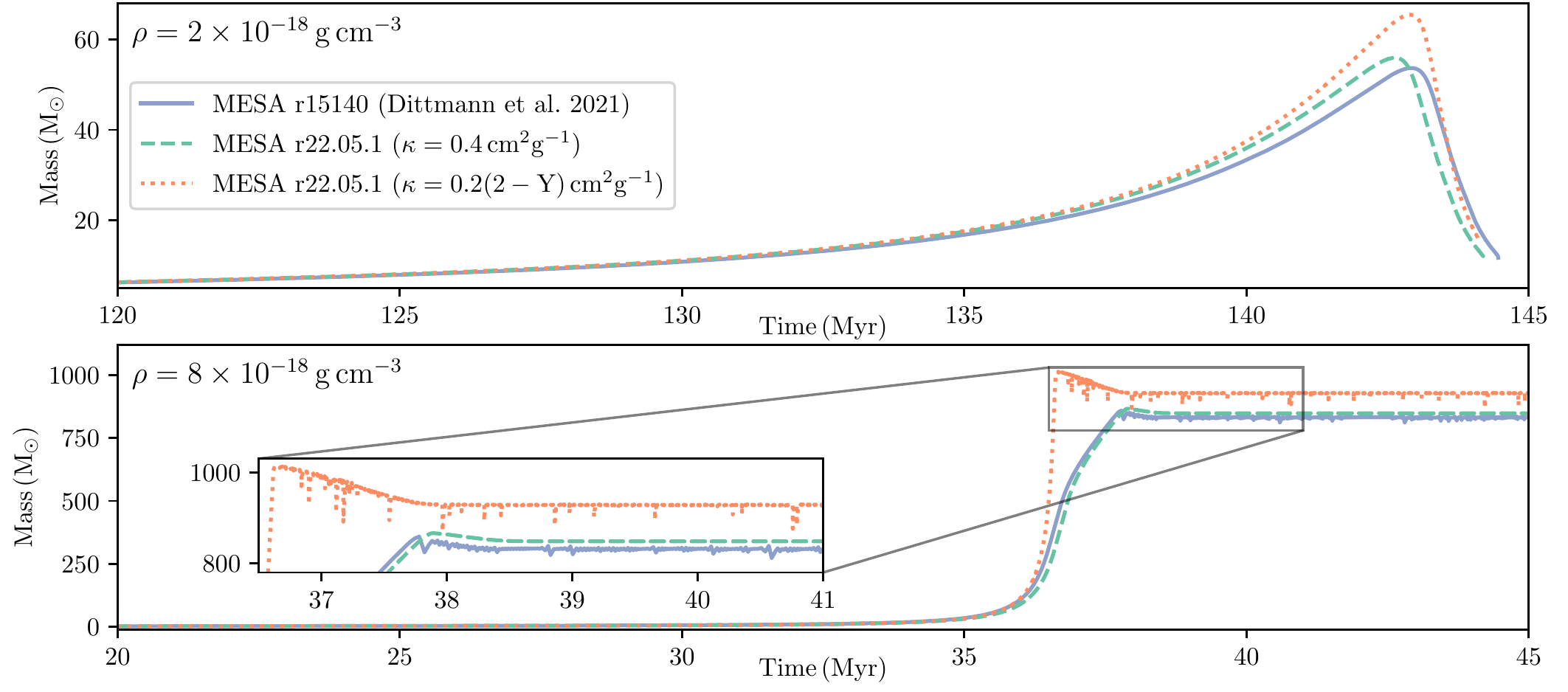}
\caption{Profiles of mass over time for fiducial immortal and intermediate stellar models illustrating differences between the models in this work and in \citet{2021ApJ...916...48D}, in all cases assuming an ambient sound speed of $c_s=10^6\,\rm{cm\,s^{-1}}$. The top panel displays models accreting gas with a density of $2\times10^{-18}\,\rm{g\,cm^{-3}}$ and the bottom panel displays models accreting gas with a density of $8\times10^{-18}\,\rm{g\,cm^{-3}}$. The solid blue lines plot models from Figures 2 and 3 of \citet{2021ApJ...916...48D}, which used \texttt{MESA} revision 15140 and assumed a constant opacity of $\kappa=0.4\,\rm{cm^2g^{-1}}$ when calculating $L_{\rm Edd}$. The dashed green lines show models using the same opacity when calculating $L_{\rm Edd}$, but are computed using \texttt{MESA} revision r22.05.1. The dotted orange lines display models calculated using \texttt{MESA} r22.05.1 which used $\kappa=0.2(2-Y)\,\rm{cm^2g^{-1}}$ when calculating $L_{\rm Edd}$. All models accreted gas with $X=0.72$ and $Y=0.28$.}
\label{fig:compare}
\end{figure*}

Of course, such stars cannot truly be immortal: the stars may migrate through the disk to regions where the accretion timescale is appreciably longer or the disk may dissipate, in which case the evolution of immortal stars proceeds similarly to intermediate models, losing most of their accreted mass and reaching later stages of burning \citep[see, e.g. Figures 10 and 11 in][]{2021ApJ...910...94C}. Additionally, we note that if the accretion timescale is shorter than the Kelvin-Helmholtz timescale for massive stars \citep[$\tau_{\rm{KH}}\sim3\times10^4\,\rm{yr,}$][]{1984ApJ...280..825B},
the luminosity of the star may not be able to adjust quickly enough to drive the star towards an equilibrium. Such stars may accrete all of the local disk mass \citep[e.g.][]{2004ApJ...608..108G}, and potentially reach the general relativistic limit of stellar stability \citep{1964ApJ...140..434T,1973ApJ...183..637B}, although our models cannot probe this regime numerically.

\section{Numerical Methods}\label{sec:numerical}
We model the evolution of AGN stars using revision r22.05.1 of the Modules for Experiments in Stellar Astrophysics \citep[\texttt{MESA};][]{{2011ApJS..192....3P},{2013ApJS..208....4P},{2015ApJS..220...15P},{2018ApJS..234...34P},{2019ApJS..243...10P},{2022arXiv220803651J}} software instrument. We implement accretion, mass loss, and modified boundary conditions following \citet{2021ApJ...910...94C}, including, e.g., treatment of effects such as the ram pressure of the accretion stream onto the star and irradiation of the star by the AGN disk. Previous simulations of stellar evolution in AGN disks \citep[e.g.][]{2021ApJ...910...94C,2021ApJ...914..105J,2021ApJ...916...48D} used \texttt{MESA} revision 15140, and it is pertinent to review some aspects of \texttt{MESA} which have changed between that version and revision r22.05.1 as they relate to our models.

The primary change between \texttt{MESA} r15140 and \texttt{MESA} r22.05.1 relevant to our calculations is the introduction of a time-dependent convection scheme, which explicitly models the growth and decay of turbulent kinetic energy in convection zones \citep{1986A&A...160..116K,2022arXiv220803651J}.
This helps our models avoid sudden `jumps' in the conditions in the outer envelope and atmospheric boundary conditions. Additionally, the default equation of state (EOS) in \texttt{MESA} r22.05.1 makes use of FreeEOS \citep{Irwin2004} and Skye \citep{2021ApJ...913...72J} in regions previously covered by OPAL \citep{Rogers2002} and PC \citep{Potekhin2010}.
The new EOS provides better thermodynamic consistency and more accurate partial derivatives, allowing the \texttt{MESA} solver to converge faster and eliminating a potential source of numerical challenges.

As in previous works, we relax the atmospheric boundary conditions and accretion rate, sequentially such that the atmospheric boundary conditions are fully relaxed before accretion begins, over a period of approximately $10^7$ years \citep[e.g.][]{2021ApJ...910...94C,2021ApJ...916...48D}. The total relaxation time is sufficiently short that it amounts to only a small fraction of the total evolutionary time of our initial $1\,M_\odot$ models. Additionally, as in previous work, we assume an enhancement of compositional mixing in radiative regions as models approach the Eddington luminosity \citep[e.g.]{2021ApJ...910...94C}.

Unlike previous studies, which fixed the composition of accreting material to $X=0.72,$ $Y=0.28$, and $Z=0$, we vary $X$ and $Y$ from $0.2$ to $0.8$ while holding $Z=0$. To prevent numerical instabilities stemming from significantly mismatched opacities between the stellar atmosphere and accretion stream at early times (particularly for models with large $Y$), we relax the initial stellar compositions to have the same $Y$ as the accreting material (holding the initial stellar $Z$ constant),\footnote{This relaxation should not affect the longer-term evolution of our models, at least for intermediate and immortal stars the mass of which is quickly dominated by accreted material. On the other hand, this relaxation procedure shortens the main sequence lifetimes of stars which do not accrete appreciably.} whereas previous studies used models with initially solar compositions \citep[e.g.][]{2021ApJ...910...94C,2021ApJ...916...48D}. 
Relaxing the composition of AGN stars to match that of the accreting material is appropriate for stars formed in the disk rather than those captured into it, although the departures from the evolution of stars with unrelaxed compositions due to this procedure are typically minor.

Because changes in $Y$ affect both stellar luminosities through $\mu$ and gas opacities through $\kappa$, we have attempted to understand both effects: while varying $Y$, we analyze two sets of simulations, one which varies only the stellar and accreting compositions while holding $\kappa=0.4\,\rm{cm^2g^{-1}}$ when evaluating $L_{\rm Edd}$, and another also adjusting the opacity used in calculating $L_{\rm Edd}$ according to $\kappa=0.2(2-Y)\,\rm{cm^2g^{-1}}$, where $Y$ is specifically the helium mass fraction of the accreting material. 

We illustrate in Figure \ref{fig:compare} the salient changes in our stellar models due to the choices made in this work compared to \citet{2021ApJ...916...48D}. 
Comparing the models shown in blue solid lines (reproduced from Figures 2 and 3 of \citet{2021ApJ...916...48D}) and the models shown in green dashed lines which make the same assumptions regarding opacity, we can observer the effects resulting from the different \texttt{MESA} versions employed. For example, the immortal model shown in the bottom panel of Figure \ref{fig:compare} shows steadier evolution with fewer ``glitches.'' We suspect this is the effect of time-dependent convection in our models, where near-surface convective zones evolve more gradually following the growth and decay of the turbulent kinetic energy.

Additionally, both models calculated using \texttt{MESA} r22.05.1 have slightly lower luminosities at a given mass than the comparable r15140 models, resulting in a slightly higher equilibrium mass in the immortal case and a slightly larger peak mass in the intermediate case. 
We note that due to relaxing the initial compositions of the models calculated in this work to match the accreting composition, those models have a slightly larger initial $\mu$ than the models from \citet{2021ApJ...916...48D}, which should increase their luminosities. Thus, the slight difference in luminosities is despite this methodological change rather than due to it. As presaged by Section \ref{sec:analytic}, the orange dotted curve in Figure \ref{fig:compare} illustrates that when we adjust the opacity as a function of $Y$, immortal stars reach higher equilibrium masses and intermediate stars reach higher peak masses. 

\section{Results}\label{sec:results}

\begin{figure}
\includegraphics[width=\linewidth]{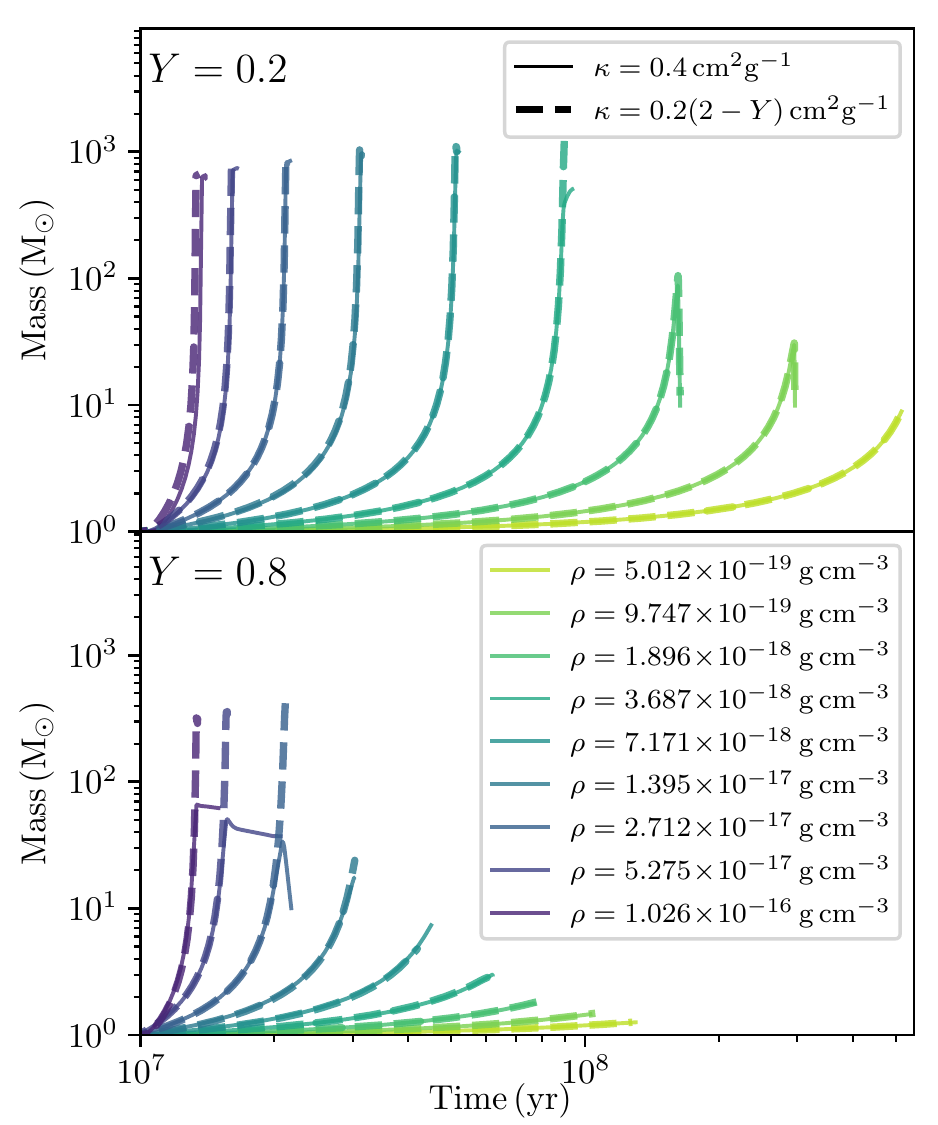}
\caption{Tracks of mass over time for a subset of the models presented in
this work, at $Y=0.2$ (top row) and $Y=0.8$ (bottom row). Calculations 
which used $\kappa=0.2(2-Y)\,\rm{cm^2\,s^{-1}}$ when computing $L_{\rm Edd}$ are shown using dashed lines, and calculations which used $\kappa=0.4\,\rm{cm^2\,g^{-1}}$ are shown using solid lines. Colors correspond to the ambient density, with darker colors corresponding to higher densities. Each simulation held $c_s=10^6\,\rm{cm\,s^{-1}}$.}
\label{fig:yRhoSweep}
\end{figure}
We have carried out a suite of AGN star simulations over a range of ambient densities and disk helium mass fractions, focusing on 15 values of $Y$ between 0.2 and 0.8 (inclusive) and densities between $\sim10^{-18}$ and $\sim10^{-16}\,\rm{g\, cm^{-3}}$, at a fixed sound speed of $c_s=10^6\,\rm{cm\,s^{-1}}$, such that the chosen range of densities covers the standard-to-immortal transition at each $Y$. From Equation (\ref{eq:bondi}), it is clear that it is $\rho c_s^{-3}$ rather than $\rho$ which governs the evolution of the AGN stars considered here, altering the initial accretion timescale. While we quote results in terms of density for the sake of brevity, it should be understood that $\rho c_s^{-3}$ is the physical quantity of import.
We carried out an initial survey over the full $(\rho-Y)$ parameter space for both of the choices of opacity employed in calculating the Eddington luminosity in this work ($\kappa=0.4\,\rm{cm^2\,g^{-1}}$ and $\kappa=0.2(2-Y)\,\rm{cm^2\,g^{-1}}$), the results from which are displayed in Figures \ref{fig:yRhoSweep} and \ref{fig:mGrid}. After identifying the general range of densities at each $Y$ and $\kappa$ where the intermediate-to-immortal transition occurs, we carried out an additional set of simulations with more fine-grained resolution in density to more precisely determine the density of the intermediate-to-immortal transition ($\rho_{\rm crit}$), the results from which are shown in Figure \ref{fig:rhoCrit}. Furthermore, we display the $^{4}$He (left column) and $^{12}$C, $^{14}$N, and $^{16}$O yields of AGN stars in Figure \ref{fig:yields}.

We find that as $Y$ increases: (1) AGN stars typically reach lower masses; (2) mortal AGN stars have shorter lifetimes (3) higher gas densities are required to sustain immortal stars; (4) AGN stars require higher densities to significantly enrich the disk with helium and metals.
\subsection{General Trends}

In order to gain a general understanding of the lives of AGN stars under different conditions, we begin by reviewing profiles of their mass over time over a range of densities at high and low Y, shown in Figure \ref{fig:yRhoSweep}. Results from simulations using $\kappa=0.4\,\rm{cm^{2}\,g^{-1}}$ and $\kappa=0.2(2-Y)\,\rm{cm^{2}\,g^{-1}}$ when calculating $L_{\rm Edd}$ are shown therein using solid and dashed lines respectively, with the former isolating changes in the evolution of AGN stars as their mean molecular weight changes with the composition of the accreted material. Regardless of our treatment of the Eddington luminosity, higher helium mass fractions of the accreted material lead to less massive immortal stars, higher densities being required to sustain immortal stars, and to intermediate stars having lower maximum masses, all following from stars being more luminous at a given mass as their mean molecular weight increases. Additionally, it appears in both cases that stars accreting higher-$Y$ material have shorter main-sequence lifetimes, although this is primarily the result of our choice to relax the initial composition.

When the Eddington luminosity increases along with the helium mass fraction of the accreting gas, the critical density for immortal stars moves to lower densities, although it still increases with $Y$. This is because the increase in the Eddington luminosity is able to partially, but not completely, counteract the changes in stellar luminosity as $Y$ increases. For the same reason, although in both cases the peak masses of immortal and intermediate stars decrease as $Y$ increases, they do so to a lesser extent when the Eddington luminosity also increases with $Y$. 

\begin{figure}
\includegraphics[width=\linewidth]{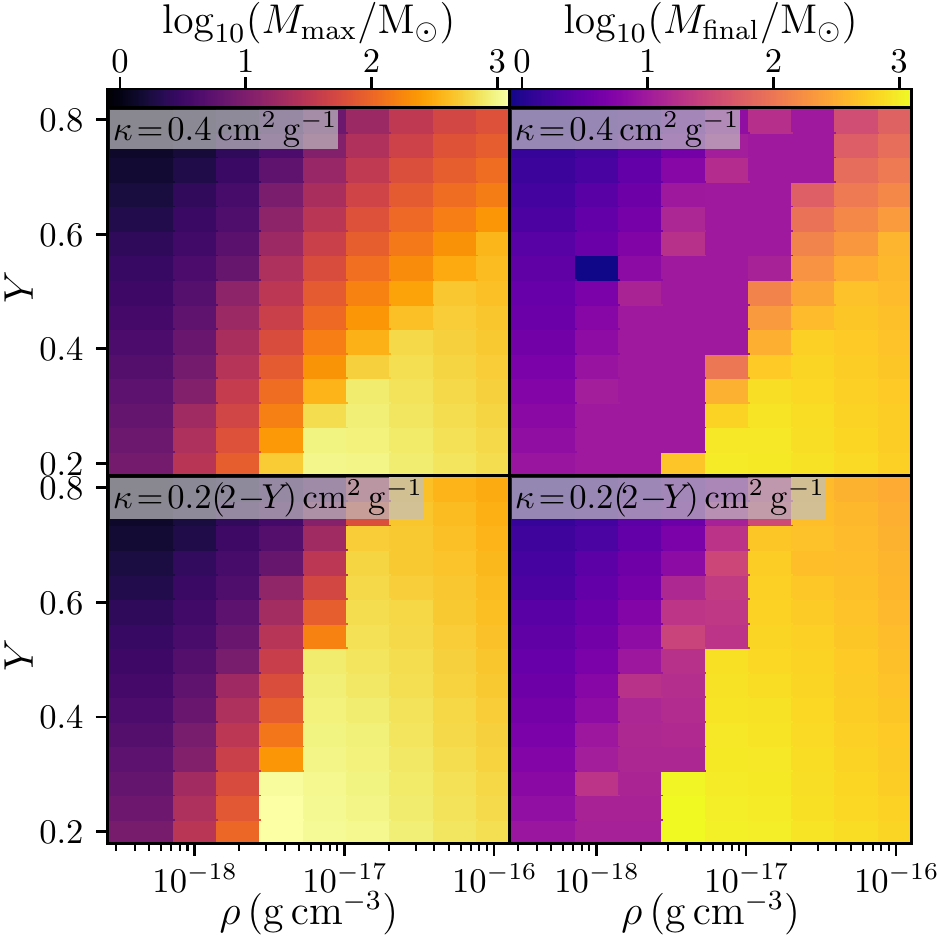}
\caption{The final (left column) and maximum (right column) masses achieved by AGN stars as functions of $\rho$ and $Y$, using both $\kappa=0.4\,\rm{cm^2\,g^{-1}}$ (top row) and $\kappa=0.2(2-Y)\,\rm{cm^2\,g^{-1}}$ when calculating $L_{\rm Edd}$. All simulations assumed an ambient sound speed of $c_s=10^6\,\rm{cm\,s^{-1}}$.}
\label{fig:mGrid}
\end{figure}

Trends in the density required to sustain immortal stars along with typical maximum and final masses (denoted $M_{\rm final}$ and $M_{\rm max}$ respectively) over the full range of densities and helium mass fractions investigated in this work are shown in Figure \ref{fig:mGrid}, in which the top row includes results from simulations which held $\kappa$ fixed when calculating $L_{\rm Edd}$, and the bottom row includes results from simulations which varied $\kappa$ in concert with $Y$. In the left column, which illustrates the maximum masses achieved by each model, we see clearly that at each constant density, as $Y$ increases stars reach smaller maximum masses, a natural result of their increased luminosity at constant mass. Similarly, for standard and intermediate stars, as well as transitioning between the intermediate and immortal regimes, maximum masses increase at constant $Y$ as density increases. 

However, considering the immortal stars alone, there is a slight trend towards lower masses as density increases, although not to as low masses as the intermediate models. We observe that these stars, which are subject to higher accretion and mass loss rates, have core metal mass fractions orders of magnitude lower than immortal stars at slightly lower ambient densities (e.g. $Z\lesssim10^{-8}$ rather than $Z\gtrsim10^{-6}$). However, these higher-ambient-density models still primarily produce energy via the CNO cycle, and have higher central densities and temperatures (e.g. $\sim10^{8.1}\,\rm K$ rather than $\sim10^{7.9}\,\rm K$). At these higher temperatures, the CNO energy generation rate is more sensitive to changes in temperature \citep{Angulo1999}, and thus these lower-Z stars develop higher luminosities at a given mass than those with slightly lower accretion rates and higher core metal mass fractions, and are thus able to reach values of $L_*/L_{\rm Edd}$ sufficient to stave off accretion at lower masses.

\begin{figure}
\includegraphics[width=\linewidth]{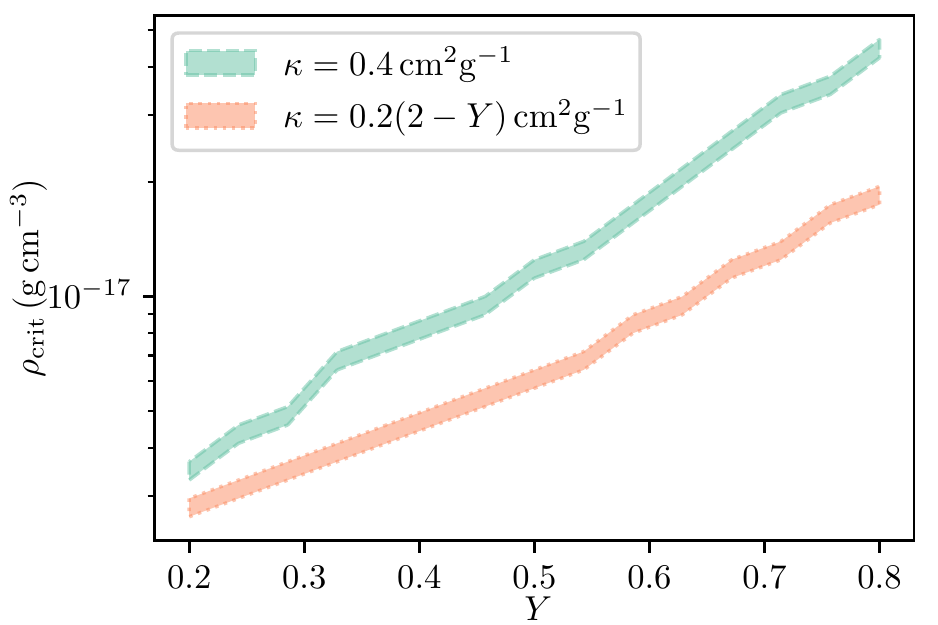}
\caption{The constraints on the density at which AGN stars transition from intermediate to immortal ($\rho_{\rm crit}$) from our simulations. The green dashed line and corresponding shaded region display results for simulations which used $\kappa=0.4\,\rm{cm^2\,g^{-1}}$ when calculating $L_{\rm Edd}$, and the orange dotted lines and corresponding shaded region display results from our simulations using $\kappa=0.2(2-Y)\,\rm{cm^2\,g^{-1}}$ when calculating $L_{\rm Edd}$. }
\label{fig:rhoCrit}
\end{figure}

The right column of Figure \ref{fig:mGrid} shows the final masses of our stellar models, and is particularly useful for differentiating immortal and intermediate stars, where the former tend to have `final' masses in excess of $100\,M_\odot$ and the latter tend to have final masses around $\sim10\,M_\odot$. 
We investigated the transition between immortal and intermediate stars more closely by performing an additional set of simulations at each $Y$ over the density range where the intermediate-to-immortal transition occurs, the results of which are shown in Figure \ref{fig:rhoCrit}. Following from the two opacity formulas differing most significantly at high $Y$ and being identical at $Y=0$, the critical densities required to support immortal stars, between the two opacity prescriptions, are closest at $Y=0.2$, gradually growing further apart as $Y$ increases. We also confirm the result gleaned from a more sparse sampling of the parameter space in Figures \ref{fig:yRhoSweep} and \ref{fig:mGrid} that, although to a lesser extent in the $\kappa=0.2(2-Y)\,\rm{cm^2\,g^{-1}}$ case, that the $\rho_{\rm crit}$ increases at higher helium mass fractions.

\begin{figure}
\includegraphics[width=\linewidth]{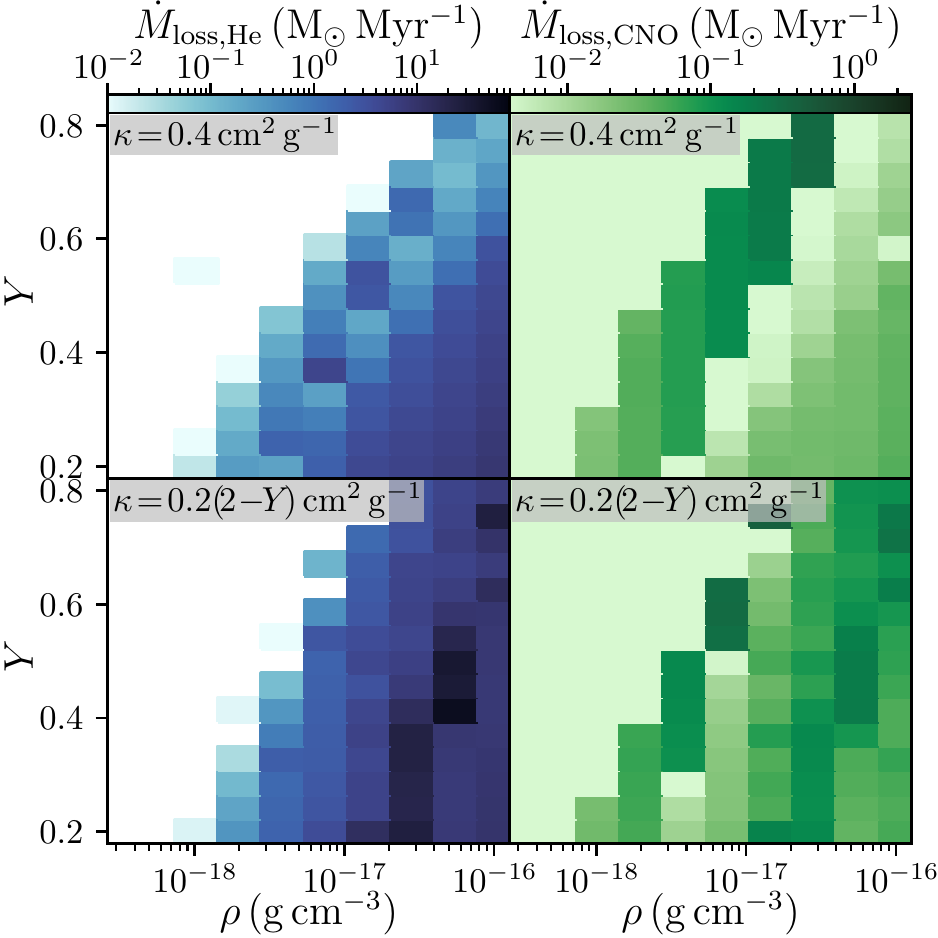}
\caption{The rate of $^{4}$He (left column) and $^{12}$C, $^{14}$N, and $^{16}$O (right column) lost from AGN stars through radiation-driven winds, varying the helium composition and density of the accreting material, holding $c_s=10^6\,{\rm cm\,s^{-1}}$. For immortal stars, these rates are averaged from the point at which the star grows to $80\%$ of its eventual maximum mass; for other stars, these rates are calculated over the full duration of each simulation. The top row shows results when only considering how the helium mass fraction of the accreting material affects the composition of the star, while the bottom row shows the yields from AGN stars when also considering how the opacity of gas varies along with its composition.
The diagonal discontinuity in the right column of the figure follows the intermediate-to-immortal transition; because intermediate stars reach later stages of nuclear burning, they can enrich the disk with metals at a higher rate than most immortal stars.} 
\label{fig:yields}
\end{figure}

\subsection{Chemical Yields}

Intermediate AGN stars, as they reach later stages of nuclear burning, expelling tens to hundreds of solar masses of material, have the potential to significantly enrich the metal content of AGN disks \citep[e.g.][]{2021ApJ...910...94C,2021ApJ...916...48D}. However, energy generation in immortal stars is dominated by the CNO cycle, and thus their winds enrich the disk with helium \citep{2021ApJ...916...48D}, which could potentially lead to AGN disks with helium mass fraction approaching unity given sufficiently numerous immortal stars \citep{2022ApJ...929..133J}. In Figure \ref{fig:yields} we illustrate how the chemical yields from AGN stars - specifically the $^4$He, $^{12}$C, $^{14}$N, and $^{16}$O content of their winds - vary with the ambient density and helium mass fractions of AGN disks. We average the mass loss rates of immortal stars starting from the point at which they first reach $80\%$ of their maximum mass, and average the mass loss rates of other stars over their entire lives.
For standard and intermediate stars, we integrate the helium lost via winds, and subtract the helium accreted over their entire lives, while for immortal stars, we report the rate of helium generated through fusion, $L_*4m_p/27 \rm{MeV},$ which follows from the difference in rest mass energy of the $^1$H and $^4$He nuclei \citep{1958ApJ...127..551F}, where $m_p$ is the proton mass. 

We observe similar trends between immortal and intermediate stars as in previous works, that immortal stars can dominate disk helium enrichment and that intermediate stars primarily enrich the disk with metals. Because intermediate stars reach later stages of burning, while immortal stars are supported primarily through the CNO cycle, intermediate stars are able to enrich the disk with metals at a much higher rate than immortal stars with lower mass loss rates, the wind metal content of which is dominated by the trace metals from CNO burning.
Thus the primary influence of composition on disk chemical enrichment follows from Figure \ref{fig:rhoCrit}, that as $Y$ increases, larger ambient densities (at constant sound speed) are required to facilitate immortal stars. 

As shown in Figures \ref{fig:yRhoSweep} and \ref{fig:mGrid}, stellar masses tend to be larger at constant density and helium mass fraction if the Eddington luminosity increases with $Y$. This naturally leads to larger rates of disk chemical enrichment, as illustrated in Figure \ref{fig:yields}. The \textit{rate} of disk metal enrichment due to intermediate stars also increases with $Y$, although this is primarily the result of shorter stellar lifetimes. We note that as intermediate stars reach later stages of nuclear burning, the metal content of their winds is dominated by carbon and oxygen, while the metal content of the winds of immortal stars is enriched primarily in nitrogen, following from their burning being dominated by the CNO cycle \citep[see, e.g. Figure 4 of][]{2021ApJ...916...48D}. Thus, the CNO yields of immortal stars shown in the right column of Figure \ref{fig:yields}, although subdominant to the mass lost in the form of helium, could significantly enhance the nitrogen content of the disks compared to the disk metal content as a whole. We also note that we have only included chemical yields from stellar winds, and that supernovae from immortal stars would further enrich AGN disks with metals.

\section{Discussion}\label{sec:discussion}
Previous works \citep[e.g.][]{2021ApJ...910...94C,2021ApJ...914..105J,2021ApJ...916...48D,2022ApJ...929..133J} have reviewed possible observational signatures of stellar evolution in AGN disks, including enrichment of the disk with helium and metals, explosive transients, gravitational wave events, and the stellar population of the Galactic Center.
Here we focus specifically on only those aspects which are directly related to disk composition.

The key way in which disk composition affects various observables is illustrated in Figure \ref{fig:rhoCrit}: that as the disk helium mass fraction increases, higher values of $\rho c_s^{-3}$ are required to produce immortal stars. Ignoring other complications such as migration through the disk, as immortal stars increase the helium mass fraction in the disk, it becomes more difficult to sustain immortal stars, which could bolster metal enrichment due to AGN stars and limit disk helium mass fractions, precluding the helium mass fractions of order unity predicted by \citet{2022ApJ...929..133J} for longer-lived disks with large populations of immortal stars. 

If immortal stellar masses also decrease sharply with increasing $Y,$ as in our calculations assuming $\kappa=0.4\,\rm{cm^2\,g^{-1}}$, the efficacy of each individual immortal star at enriching the disk with helium would decrease further with increasing $Y,$ as shown in Figure \ref{fig:yields}. However, if the opacity decreases with $Y$ then immortal stellar masses decrease at a much slower rate as $Y$ increases, and are generally larger, supporting higher rates of helium enrichment.
Whether or not there is a limiting value of $Y$ for a given disk due to enrichment by immortal stars depends on disk density, sound speed, the accretion rate through the disk, and the number of embedded immortal stars. However, higher values of $Y$ narrow the viable range of disk parameters where immortal stars can produce a flux of new helium into the disk comparable to the accretion rate through the disk.

It is likely that the Milky Way experienced a period of nuclear activity approximately 2-8 Myr ago \citep[e.g.][]{Su:2010,Bland-Hawthorn:2019}. Contrary to theoretical expectations \citep{Bahcall:1976,Bahcall:1977}, the fraction of low-mass stars decreases moving towards the galactic center \citep{Genzel:2010,Do:2017}, which contains an unexpectedly large fraction of young massive stars \citep{Ghez:2003, Levin:2003, Alexander:2005,2006ApJ...643.1011P}. Such unusual stars may have formed in \citep[e.g.][]{2007MNRAS.374..515L} or been captured into and rejuvenated by \citep[e.g.][]{2020MNRAS.498.3452D,2021ApJ...910...94C} the gas disk present during the most recent active phase of the milky way. Additionally, spectroscopic observations suggest that some of the stars in the Galactic Center may be enriched with helium \citep{Martins:2008,Habibi:2017,Do:2018}, which would follow naturally from formation in or accretion from a helium-enriched disk. 

Observations suggest that many AGN may have helium mass fractions $Y\gtrsim0.3$, indicative of significant helium enrichment \citep{1971ApJ...163..235B,1975ApJ...201...26B,2022MNRAS.tmp.1674D}. The highest of the
narrow-line region measurement presented in \citet{2022MNRAS.tmp.1674D} is $Y\sim0.46$, rather than the near-unity values predicted for long-lived disks with large populations of immortal stars \citep{2022ApJ...929..133J}. However, the narrow-line region would be enriched primarily through AGN-driven winds and likely possesses a helium mass fraction lower than the broad-line region and AGN disk as a whole. Mass-loss from embedded stars should lead to enhanced broad-line-region helium abundances relative to those in the narrow-line region once these can be reliably measured, although detailed predictions of these effects (as well as the rates of gravitational wave events and electromagnetic transients) require incorporating the effects of disk composition, as well as those of the local density, sound speed, and tidal forces \citep{2021ApJ...910...94C,2021ApJ...916...48D} into more detailed models along the lines of but extending \citet{2022ApJ...928..191G,2022arXiv220510382D}.

A subset of $\sim 1\%$ of quasars are nitrogen-rich, displaying anomalously strong nitrogen lines and elevated nitrogen-to-carbon abundance ratios \citep[e.g.][]{{2004AJ....128..561B},{2004AJ....127..576B},{2008ApJ...679..962J},{2014MNRAS.439..771B},{2009A&A...503..721M}}. Decreases in the nitrogen-to-carbon line ratios of these sources over time have been linked to tidal disruptions in some AGNs \citep{{2016MNRAS.458..127K},{2018ApJ...859....8L}}, although line ratios in some nitrogen-rich quasars can remain constant or grow over time \citep{2018ApJ...859....8L}. Preferential nitrogen enrichment from the winds of immortal stars may contribute to the high nitrogen abundances observed in these sources, particularly in AGN with SMBHs too massive to tidally disrupt stars.

\section{Conclusions}\label{sec:conclusion}
The helium content of AGN disks can play a significant role in determining the evolution of stars which form in or are captured into the disk. Observations suggest that some quasars may have helium mass fractions as high as $Y\sim0.5$ \citep{1975ApJ...201...26B,2022MNRAS.tmp.1674D}; theoretical predictions suggest that embedded stellar populations may drive the helium content of AGN disk towards unity \citep{2022ApJ...929..133J}; and massive stars in the Galactic Center, potentially remnants from a previous accretion episode onto Sagittarius A$^*$\citep[e.g.][]{2007MNRAS.374..515L}, show signs of helium enrichment \citep{Do:2018}. Accordingly, we studied the role of the helium mass fraction of AGN disks on the evolution of stars embedded therein over a range of ambient densities, and thus accretion timescales. 

We found that order of magnitude higher densities can be required to sustain immortal stars in $Y=0.8$ disks compared to $Y=0.2$ disks, and the maximum masses of AGN stars tend to be lower in higher-$Y$ disks. These result from AGN stars having higher luminosities at a given mass as their mean molecular weights increase, following from their increased helium mass fractions. These higher luminosities are able to reduce the accretion rate onto AGN stars and drive mass loss at smaller masses, but can be partially counteracted if the disk opacity decreases with increasing $Y$. Both the reduction of stellar masses with increasing $Y$ and the increased difficulty in sustaining immortal stars could limit disk helium mass fraction and the immortal stellar population in AGN disks, and should be taken into account in future studies of the interaction between stars and AGN disks. 

\section*{Software}
\texttt{MESA} \citep[][\url{http://MESA.sourceforge.net}]{{2011ApJS..192....3P},{2013ApJS..208....4P},{2015ApJS..220...15P},{2018ApJS..234...34P},{2019ApJS..243...10P},{2022arXiv220803651J}},
\texttt{MESASDK} \citep{mesasdk_linux},
\texttt{matplotlib} \citep{4160265}, \texttt{numpy} \citep{5725236}

\section*{Acknowledgments}
We thank Cole Miller and Sylvain Veilleux for helpful suggestions and conversations.
Computations were performed using the Rusty cluster of the Flatiron Institute and the YORP cluster administered by the Center for Theory and Computation within the Department of Astronomy at the University of Maryland. The Center for Computational Astrophysics at the Flatiron Institute is supported by the Simons Foundation. A.J.D. was supported in part by NASA ADAP grant 80NSSC21K0649, and by LANL/LDRD under project number 20220087DR. The LA-UR number is LA-UR-22-29050.
\appendix

\section{Software details}\label{apndx:A}
Calculations were carried out using \texttt{MESA} version r22.05.1

The \texttt{MESA} EOS is a blend of the OPAL \citep{Rogers2002}, SCVH
\citep{Saumon1995}, FreeEOS \citep{Irwin2004}, HELM \citep{Timmes2000},
PC \citep{Potekhin2010}, and Skye \citep{2021ApJ...913...72J} EOSes.

Radiative opacities are primarily from OPAL \citep{Iglesias1993,
Iglesias1996}, with low-temperature data from \citet{Ferguson2005}
and the high-temperature, Compton-scattering dominated regime by
\citet{Poutanen2017}.  Electron conduction opacities are from
\citet{Cassisi2007}.

Nuclear reaction rates are from JINA REACLIB \citep{Cyburt2010}, NACRE \citep{Angulo1999} and
additional tabulated weak reaction rates \citet{Fuller1985, Oda1994,
Langanke2000}.  Screening is included via the prescription of \citet{Chugunov2007}.
Thermal neutrino loss rates are from \citet{Itoh1996}.

We adopted a 21-isotope nuclear network (approx21.net).
We used the Schwarzschild criterion to determine convective
boundaries and did not include convective overshooting.

\bibliographystyle{aasjournal}
\bibliography{references}

\end{document}